\documentclass[twocolumn,english,prl,priprint,twocolumn,superscriptaddress]{revtex4-1}
\usepackage[T1]{fontenc}
\usepackage[latin9]{inputenc}
\usepackage{geometry}
\usepackage{amsmath}
\usepackage{amssymb}
\usepackage{graphicx}

\makeatletter

\usepackage{babel}

\makeatother

\usepackage{babel}
\begin{document}
\title{Theory of anomalous Hall effect in transition-metal pentatellurides
$\mathrm{ZrTe}_{5}$ and $\mathrm{HfTe}_{5}$}
\author{Huan-Wen Wang}
\affiliation{School of Physics, University of Electronic Science and Technology
of China, Chengdu 611731, China}
\affiliation{Department of Physics, The University of Hong Kong, Pokfulam Road,
Hong Kong, China}
\author{Bo Fu}
\affiliation{Department of Physics, The University of Hong Kong, Pokfulam Road,
Hong Kong, China}
\affiliation{School of Sciences, Great Bay University, Dongguan 523000, China}
\author{Shun-Qing Shen}
\email{sshen@hku.hk}

\affiliation{Department of Physics, The University of Hong Kong, Pokfulam Road,
Hong Kong, China}
\date{\today}
\begin{abstract}
The anomalous Hall effect has considerable impact on the progress
of condensed matter physics and occurs in systems with time-reversal
symmetry breaking. Here we theoretically investigate the anomalous
Hall effect in nonmagnetic transition-metal pentatellurides $\mathrm{ZrTe_{5}}$
and $\mathrm{HfTe}_{5}$. In the presence of Zeeman splitting and
Dirac mass, there is an intrinsic anomalous Hall conductivity induced
by the Berry curvature in the semiclassical treatment. In a finite
magnetic field, the anomalous Hall conductivity rapidly decays to
zero for constant spin-splitting and vanishes for the magnetic-field-dependent
Zeeman energy. A semiclassical formula is derived to depict the magnetic
field dependence of the Hall conductivity, which is beneficial for
experimental data analysis. Lastly, when the chemical potential is
fixed in the magnetic field, a Hall conductivity plateau arises, which
may account for the observed anomalous Hall effect in experiments.
\end{abstract}
\maketitle

\section{Introduction}

The transition-metal pentatellurides $\mathrm{ZrTe}_{5}$ and $\mathrm{HfTe}_{5}$
are prototypes of massive Dirac materials with finite band gap, which
are very close to the topological transition point \citep{weng2014prx,Chen2015prl,Li16natphys,zhang2017nc,zhang2017scibull,mutch2019evidence,zhang2021natcomm,tang2019nature,jiang2020prl}.
Further studies uncover more exotic physics in these compounds, such
as the quantum anomaly \citep{Li16natphys,wang2021prb}, three-dimensional
quantum Hall effect \citep{tang2019nature,wang2020prb,Galeski2020nc,Galeski2021nc,qin2020prl,Gooth2022rpp,wang2023prb},
resistivity anomaly \citep{Okada1980jpsj,Izumi1981ssc,Izumi1982,Tritt1999prb,rubinstein1999hfte,shahi2018prx}
and anomalous Hall effect \citep{liang2018NP,sun2020npj,liu2021PRB,Mutch2021arxiv,Lozano2022arxiv,Gourgout2022arxiv,Choi2020prb}.
The anomalous Hall effect refers to the Hall effect in the absence
of an external magnetic field which typically occurs in magnetic solids
with broken time-reversal symmetry \citep{xiao2010rmp,nagaosa2010rmp}.
When an external field is applied, due to the lack of convincing calculations
based on the microscopic model, the analyses often rely on an empirical
relation \citep{nagaosa2010rmp}. In the empirical formula, the anomalous
part of Hall conductivity $\sigma_{xy}^{A}=\sigma_{0}^{A}\tanh(B/B_{0})$
reaches saturation at $\sigma_{0}^{A}$ in a large magnetic field
$B\gg B_{0}$. $\mathrm{ZrTe_{5}}$ and $\mathrm{HfTe_{5}}$ are nonmagnetic
topological materials without the prerequisite for anomalous Hall
effect at zero field, but the Hall conductivities are still found
to saturate in several tesla in experiments. Therefore, the physical
origin of the anomalous Hall effect therein is still under debate.
In systems with resistivity anomaly, the anomalous Hall effect can
be explained by the Dirac polaron picture at high temperature \citep{fu2020prl,wang2021prl}.
However, this picture cannot explain the nonlinear Hall resistivity
at low temperatures, where the temperature effect becomes unimportant
as $T\to0$, the thermal excitation of electrons from valance band
to conduction band is suppressed. In such a case, there are several
mechanisms that have been discussed in literatures. First, the multi-band
model is one possible mechanism. However, as revealed by the angle-resolved
photoemission spectroscopy (ARPES) measurement, there is only one
Fermi pocket near the $\Gamma$ point in $\mathrm{ZrTe_{5}}$, eliminating
the possibility of a multi-band effect at low temperatures. The second
viewpoint is the Zeeman effect induced Weyl nodes for massless Dirac
fermion \citep{liang2018NP,liu2021PRB,Choi2020prb}, where the induced
anomalous Hall effect is proportional to the distance of two Weyl
nodes \citep{burkov2011prl,zyuzin2012prb}. Another scenario involves
finite Berry curvature in spin-split massive Dirac fermions \citep{liu2021PRB,Mutch2021arxiv,Lozano2022arxiv}.
In semiclassical theory, a strong magnetic field is required to obtain
a sizable anomalous Hall effect, ensuring that the energy bands of
different spins are well-separated. However, when the magnetic field
is strong, the semiclassical description of the anomalous Hall effect
might be invalid. The existing discussion should be revised in a quantum
mechanical formalism.

In this work, we begin with the massive Dirac fermion with Zeeman
splitting, and investigate the Hall conductivity in it. To treat the
anomalous Hall effect and the conventional orbital Hall effect on
an equal footing, the Landau levels in a finite magnetic field are
considered. When $B\to0$, the Kubo formula gives the anomalous Hall
conductivity in the semiclassical theory for a constant spin splitting.
However, when the band broadening is much smaller than the Landau
band spacing in the strong magnetic field, the anomalous Hall conductivity
decays to zero very quickly. Based on the numerical results, we propose
a simple semiclassical equation for the total Hall conductivity from
the electrons' equation of motion, which captures the function behavior
of Hall conductivity from the weak magnetic field to strong magnetic
field very well. For the magnetic-field-dependent Zeeman splitting,
it is hard to see any signals of anomalous Hall effect from the total
Hall conductivity. Hence, the Zeeman effect is excluded as an explanation
for the anomalous Hall effect in $\mathrm{ZrTe_{5}}$. If the chemical
potential is fixed in the magnetic field due to the localization effect,
a plateau structure is observed in the Hall conductivity, which could
provide an explanation for the observed anomalous Hall effect in experiments.

\section{Model Hamiltonian and band structure}

In a finite magnetic field, the low energy Hamiltonian for $\mathrm{ZrTe}_{5}$
can be described by the anisotropic massive Dirac equation as \citep{Chen2015prl,tang2019nature,jiang2020prl}
\begin{align}
H(k)= & m\tau_{z}+\omega\sigma_{z}+\sum_{i=x,y,z}v_{i}\Pi_{i}\Gamma_{i}\label{eq:massive dirac}
\end{align}
where $\Gamma_{1}=\tau_{x}\sigma_{z}$, $\Gamma_{2}=\tau_{y}$, $\Gamma_{3}=\tau_{x}\sigma_{x}$,
$\sigma$ and $\tau$ are the Pauli matrices acting on the spin and
orbit space, respectively. $v_{i}$ with $i=x,y,z$ are the fermi
velocities along $i$-direction, $\Pi_{i}=\hbar k_{i}+eA_{i}$ are
the kinematic momentum operators and $\hbar k_{i}$ are the momentum
operators. $2m$ is the Dirac band gap, and $\omega$ is the term
related to the Zeeman splitting. For a perpendicular magnetic field,
the gauge potential can be chosen as $\mathbf{A}=(-By,0,0)$. By introducing
the ladder operators $a=\frac{(v_{x}\Pi_{x}-iv_{y}\Pi_{y})}{\sqrt{2e\hbar Bv_{x}v_{y}}}$
and $a^{\dagger}=\frac{(v_{x}\Pi_{x}+iv_{y}\Pi_{y})}{\sqrt{2e\hbar Bv_{x}v_{y}}}$
\citep{shen05prb}, the energy spectrum of Landau levels can be solved
as (see Appendix B for details) 
\begin{align}
\varepsilon_{n\zeta s} & =\zeta\sqrt{(E_{n}+s\omega)^{2}+\left(v_{z}\hbar k_{z}\right)^{2}}
\end{align}
where $E_{n}=\sqrt{m^{2}+n\eta^{2}}$, $s=\pm$ represents two splitting
states because of the Zeeman effect for $n>0$, and $E_{n}=-m,s=+$
for $n=0$. $\zeta=+$ is for the conduction band and $\zeta=-$ is
for the valence band, $\eta=\sqrt{2v_{x}v_{y}}\hbar/\ell_{B}$ is
the cyclotron energy, $\ell_{B}=\sqrt{\hbar/eB}$ is the magnetic
length. Without loss of generality, we choose the model parameters
as $m=5\,\mathrm{meV},$ $v_{x}=6.85\times10^{5}\mathrm{m/s}$, $v_{y}=4.1\times10^{5}\mathrm{m/s}$,
$v_{z}=5\times10^{4}\mathrm{m/s}$ according to Ref. \citep{jiang2020prl}.

\section{Hall conductivity in finite magnetic fields}

In the semiclassical theory, the intrinsic anomalous Hall effect can
be attributed to the nonzero Berry curvature induced by the Zeeman
effect. The obtained anomalous Hall effect is odd in the Zeeman energy
$\text{\ensuremath{\omega} }$ and band gap $2m$ (see more details
in Appendix A). In a finite magnetic field, besides the intrinsic
anomalous Hall effect at $B=0$, the orbital contribution from the
Drude formula $\sigma_{xy}^{N}\sim\frac{\chi\sigma_{D}B}{1+\chi^{2}B^{2}}$
should also be important, where $\chi$ is the electric mobility,
$\sigma_{D}$ is the zero field Drude conductivity \citep{Pippard1989book}.
Hence, we need to treat the two parts on an equal footing. The total
Hall conductivity for a disordered system can be evaluated by the
Kubo-Streda formula \citep{Mahan,Streda1982,wang2018prb} 
\begin{align}
\sigma_{xy}= & \mathrm{Im}\frac{e^{2}\hbar}{\pi V}\sum_{k}\int_{-\infty}^{+\infty}n_{F}(\epsilon-\mu)d\epsilon\nonumber \\
 & \times\text{Tr}[\hat{v}^{x}\frac{dG^{R}}{d\epsilon}\hat{v}^{y}\mathrm{Im}G^{R}-\hat{v}^{x}\mathrm{Im}G^{R}\hat{v}^{y}\frac{dG^{A}}{d\epsilon}],\label{eq:kubo-streda}
\end{align}
where $G^{R/A}=[\epsilon-H\pm i\gamma]^{-1}$ is the retarded or advanced
Green's function, $\gamma$ is the disorder induced band broadening,
$\hat{v}^{x}=i\hbar^{-1}[H,x]$ and $\hat{v}^{y}=i\hbar^{-1}[H,y]$
are the velocity operators along the $x-$ and $y-$ direction, respectively.
$n_{F}(\epsilon-\mu)=[1+\exp(\frac{\epsilon-\mu}{k_{B}T})]^{-1}$
is the Fermi-Dirac distribution function with $\mu$ the chemical
potential and $k_{B}T$ the product of Boltzmann constant and absolute
temperature. Kubo-Streda formula already includes the anomalous Hall
conductivity and orbital Hall conductivity simultaneously. To understand
the effect of Zeeman splitting on the anomalous Hall conductivity,
we study two typical cases, i.e., the constant spin-splitting and
the magnetic-field-dependent Zeeman splitting based on Eq. (\ref{eq:kubo-streda}).

\subsection{Clean limit}

To compare with intrinsic contribution in the semiclassical theory
, we first focus on the Hall conductivity in the disorder-free case,
where the Hall conductivity in the Landau level basis can be evaluated
as (see Appendix B for details)
\begin{align}
\sigma_{xy}= & -\frac{e^{2}\eta^{2}}{2\pi v_{x}v_{y}\hbar}\int_{-\infty}^{+\infty}\frac{dk_{z}}{2\pi}\sum_{\lambda\lambda^{\prime}}[v_{\lambda\lambda^{\prime}}^{(1)}]^{2}\delta_{n,n^{\prime}-1}\nonumber \\
 & \times\frac{n_{F}(\varepsilon_{\lambda}-\mu)-n_{F}(\varepsilon_{\lambda^{\prime}}-\mu)}{(\varepsilon_{\lambda}-\varepsilon_{\lambda^{\prime}})^{2}},\label{eq:Hall-clean}
\end{align}
where the subscript $\lambda$ denotes quantum numbers $\zeta,s,n$.
The product of matrix elements of $\hat{v}^{x}$ and $\hat{v}^{y}$
satisfies $v_{\lambda\lambda^{\prime}}^{x}v_{\lambda^{\prime}\lambda}^{y}=-i[v_{\lambda\lambda^{\prime}}^{(1)}]^{2}\delta_{n,n^{\prime}-1}+i[v_{\lambda^{\prime}\lambda}^{(1)}]^{2}\delta_{n,n^{\prime}+1}$.
To perform the summation over $\lambda$ and $\lambda^{\prime}$,
we take advantage of following relations,
\begin{equation}
\sum_{s^{\prime}\zeta^{\prime}}\left(\frac{v_{ns\zeta,n+1s^{\prime}\zeta^{\prime}}^{(1)}}{\varepsilon_{ns\zeta}-\varepsilon_{n+1s^{\prime}\zeta^{\prime}}}\right)^{2}=\frac{v_{x}v_{y}}{2\eta^{2}}(2n+1-\frac{sm}{E_{n}}),
\end{equation}
\begin{equation}
\sum_{s\zeta}\left(\frac{v_{ns\zeta,n+1s^{\prime}\zeta^{\prime}}^{(1)}}{\varepsilon_{ns\zeta}-\varepsilon_{n+1s^{\prime}\zeta^{\prime}}}\right)^{2}=\frac{v_{x}v_{y}}{2\eta^{2}}(2n+1-\frac{s^{\prime}m}{E_{n+1}}).
\end{equation}
Then,
\begin{align}
\sigma_{xy}= & -\frac{en_{0}}{B},\label{eq:Hall_LL}
\end{align}
where $n_{0}$ is the carrier density in the Landau level basis,
\begin{align}
n_{0}= & \frac{e^{2}}{4\pi^{2}\hbar}\int_{-\infty}^{+\infty}dk_{z}\sum_{\lambda}\sum_{\chi=\pm}\chi\theta(\chi\varepsilon_{\lambda})n_{F}[\chi(\varepsilon_{\lambda}-\mu)].\label{eq:carrier density B}
\end{align}
Hence, the Hall conductivity is always proportional to the carrier
density and the inverse of magnetic field. Even in the presence of
a finite Zeeman energy, the anomalous Hall effect is zero in the clean
limit regardless of the magnitude of magnetic field and temperature
once the carrier density $n_{0}$ is fixed. However, $\sigma_{xy}$
should be finite not divergent at zero-magnetic-field. Such a discrepancy
between the results from the zero magnetic field and finite magnetic
field is also found in a system without anomalous Hall effect. This
contradiction can be removed by considering a finite disorder scattering
in Eq. (\ref{eq:kubo-streda}).

\begin{figure}
\centering{}\includegraphics[width=7.5cm]{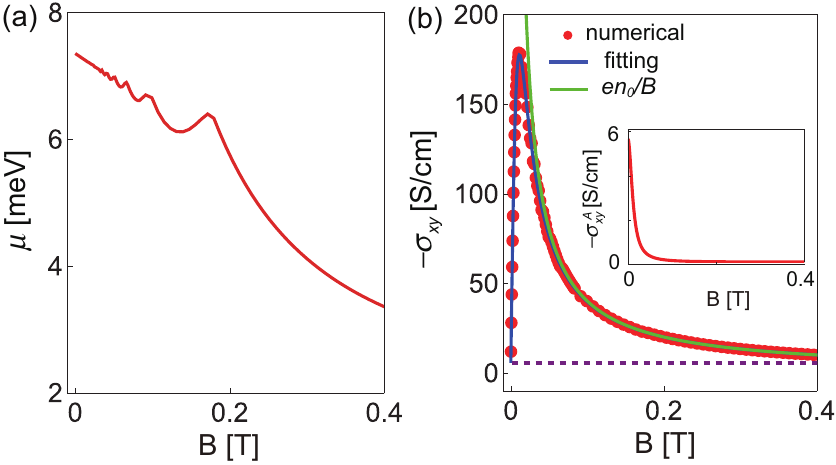}\caption{\label{fig:AHE_B_1}(a) By fixing the carrier density as $n_{0}=2.5\times10^{15}\,\mathrm{cm}^{-3}$,
the chemical potential $\mu$ is a function of magnetic field. (b)
Hall conductivity as a function of magnetic field for a constant spin
splitting $\omega=3\,\mathrm{meV}$ and constant broadening $\gamma=0.1\,\mathrm{meV}$,
the red dots are the numerical results, blue line is the fitting curve
from the Hall conductivity in Eq. (\ref{eq:empirical equation}),
the purple dashed line denotes the anomalous Hall conductivity at
zero magnetic field, and the green solid line represents the Hall
conductivity in the clean limit. The insert shows the fitted anomalous
Hall conductivity as a function of magnetic field.}
\end{figure}

\subsection{Constant spin-splitting}

\begin{figure}
\centering{}\includegraphics[width=7.5cm]{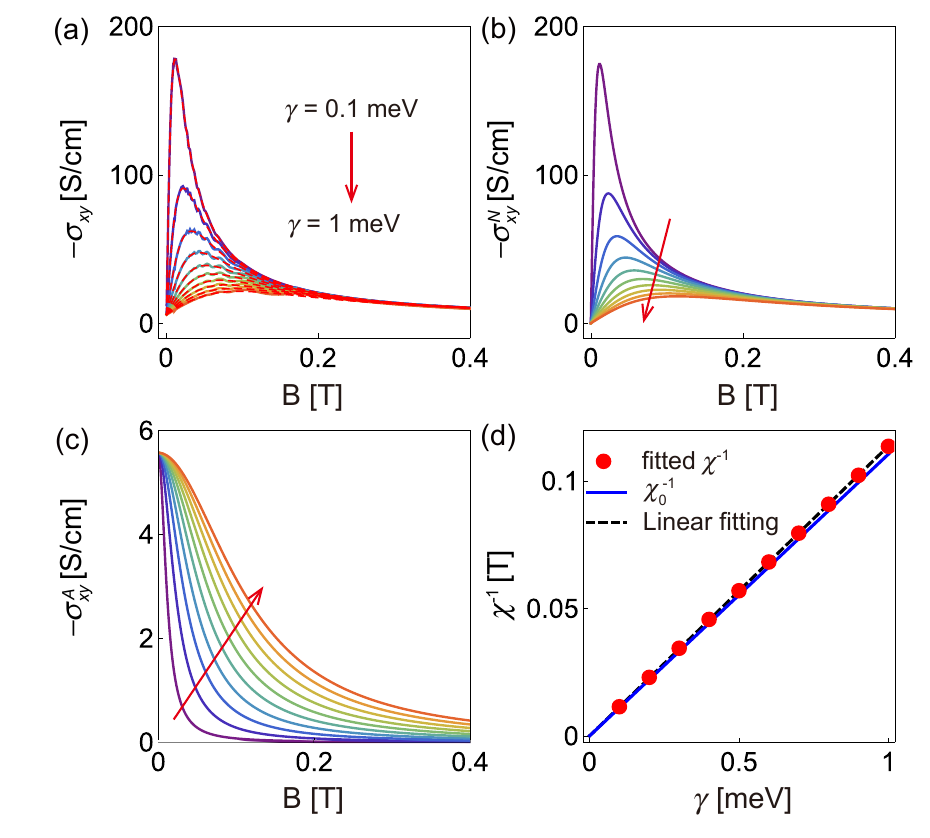}\caption{\label{fig:AHE_B_1-1}(a) Hall conductivity as a function of magnetic
field for a constant spin splitting $\omega=3\,\mathrm{meV}$ and
several selected band broadenings, the red dashed lines are the fitting
curve from Eq. (\ref{eq:empirical equation}), and the solid lines
are the numerical results from Eq. (\ref{eq:kubo-streda}). (b) and
(c) are the fitted orbital part and anomalous part of Hall conductivity
in (a). (d) the inverse of fitted mobility as function of band broadenings,
the blue lines are the inverse of mobility at $B=0$, $\chi_{0}=\frac{e\hbar}{2\gamma}\frac{v_{x}v_{y}}{(\mu+\omega)}$,
and the dashed line is the linear fitting to $\chi^{-1}$.}
\end{figure}

For a constant spin-splitting, there is a finite anomalous Hall effect
at $B=0$, and its magnitude decreases with the increasing of magnetic
field. Here we choose the calculation parameters as $\gamma=0.1\,\mathrm{meV}$,
$\omega=3\,\mathrm{meV}$, and $n_{0}=2.5\times10^{15}\,\mathrm{cm^{-3}}$.
By fixing carrier density in the magnetic field, the chemical potential
can be solved out from the definition of $n_{0}$ in Eq. (\ref{eq:carrier density B}).
As shown in Fig. \ref{fig:AHE_B_1}(a), the chemical potential decreases
linearly with increasing magnetic field in the weak magnetic field
region and oscillates with the field in the strong magnetic field
region. Plugging the chemical potential at finite magnetic field into
the Kubo-Streda formula, we obtain the Hall conductivity as indicated
by open circles in Fig. \ref{fig:AHE_B_1}(b). The Hall conductivity
approaches the numerical value of anomalous Hall effect in the zero
magnetic filed (purple dashed line). To have a quantitative description
for the field dependence of anomalous Hall conductivity, we phenomenologically introduce
the transport equation for charge carriers in the presence of electric
and magnetic fields, which takes the following form
\begin{equation}
\mathbf{j}=\sigma_{D}\mathbf{E}+\chi\mathbf{j}\times\mathbf{B}+\sigma_{A}\mathbf{E}\times\hat{\mathbf{z}}.
\end{equation}
Here $\mathbf{j}$ is the electric current density, the magnetic field
$\mathbf{B}$ is along the $z$-direction, $\sigma_{A}$ is the anomalous
Hall conductivity at $B=0$ describing the Hall response in $x-y$ plane. The second term is given by the Lorentz force experienced by charge carriers in a magnetic field. After some vector algebra, we can obtain
the field-dependent Hall conductivity as
\begin{equation}
\sigma_{xy}=\frac{\sigma_{A}+\chi B\sigma_{D}}{1+\chi^{2}B^{2}}.\label{eq:empirical equation}
\end{equation}
The denominator $1+\chi^{2}B^{2}$ indicates that the anomalous Hall
conductivity is suppressed at the high field as $\chi B\gg1$. Especially,
the anomalous Hall conductivity becomes zero in the clean limit as
$\chi\to+\infty$. As shown in Fig. \ref{fig:AHE_B_1}(b), the calculated
Hall conductivity {[}red dots{]} can be well-fitted by Eq. (\ref{eq:empirical equation})
{[}blue line{]} in the full magnetic field regime. In the insert of
Fig. \ref{fig:AHE_B_1}(b), we present the fitted anomalous Hall conductivity
$\sigma_{xy}^{A}$ as function of magnetic field, it decays to zero
very quickly in the high field. A similar magnetic field dependence
of $\sigma_{xy}^{A}$ has also been found in two dimensional systems
\citep{Tsaran2016prb}. Besides, we plot the corresponding Hall conductivity
in the clean limit ($\gamma=0$) in Fig. \ref{fig:AHE_B_1}(b) for
comparison (green solid line), where we have used the analytical expression
$\sigma_{xy}=-en_{0}/B$. In the weak magnetic field, $\chi B\to0$,
the disorder effect is prominent and removes the divergence of the
orbital part of $\sigma_{xy}$. While in a strong magnetic field,
if the energy spacing of Landau levels becomes larger than the band
broadening, one can ignore the disorder effect; then, the Hall conductivities
with and without disorder effect coincide with each other in the high
field regime. We present the Hall conductivity in a finite magnetic
field by choosing several band broadenings in Fig. \ref{fig:AHE_B_1-1}.
The background of total Hall conductivity {[}solid lines{]} can be
well-fitted by Eq. (\ref{eq:empirical equation}) as indicated by
the red dashed line in Fig. \ref{fig:AHE_B_1-1}(a). Accordingly,
we plot the fitted orbital part $\frac{\chi B\sigma_{D}}{1+\chi^{2}B^{2}}$
and anomalous part $\frac{\sigma_{A}}{1+\chi^{2}B^{2}}$ in Fig. \ref{fig:AHE_B_1-1}(b)
and (c), respectively. The orbital Hall conductivities are suppressed
in the low magnetic field by the band broadening, and collapse together
in the high magnetic field. While for the anomalous Hall conductivities,
they are almost independent of the band broadening at $B=0$, and
increase with the increasing of band broadening in a finite magnetic
field. As shown in Fig. \ref{fig:AHE_B_1-1}(d), the obtained mobility
{[}red dots{]} is inversely proportional to the band broadenings as
indicated by the dashed line. It is noted that the fitted $\chi$
is slightly larger than the mobility at zero magnetic field $\chi_{0}=\frac{e\hbar}{2\gamma}\frac{v_{x}v_{y}}{(\mu+\omega)}$,
which might be caused by the field-dependent chemical potential in
Fig. \ref{fig:AHE_B_1}(a). Hence, Eq. (\ref{eq:empirical equation})
indeed quantitatively captures the magnetic field dependence of Hall
conductivity, and the anomalous Hall effect vanishes in the high magnetic
filed and does not display a step like function.

\subsection{Magnetic-field-dependent Zeeman splitting}

\begin{figure}
\centering{}\includegraphics[width=7.5cm]{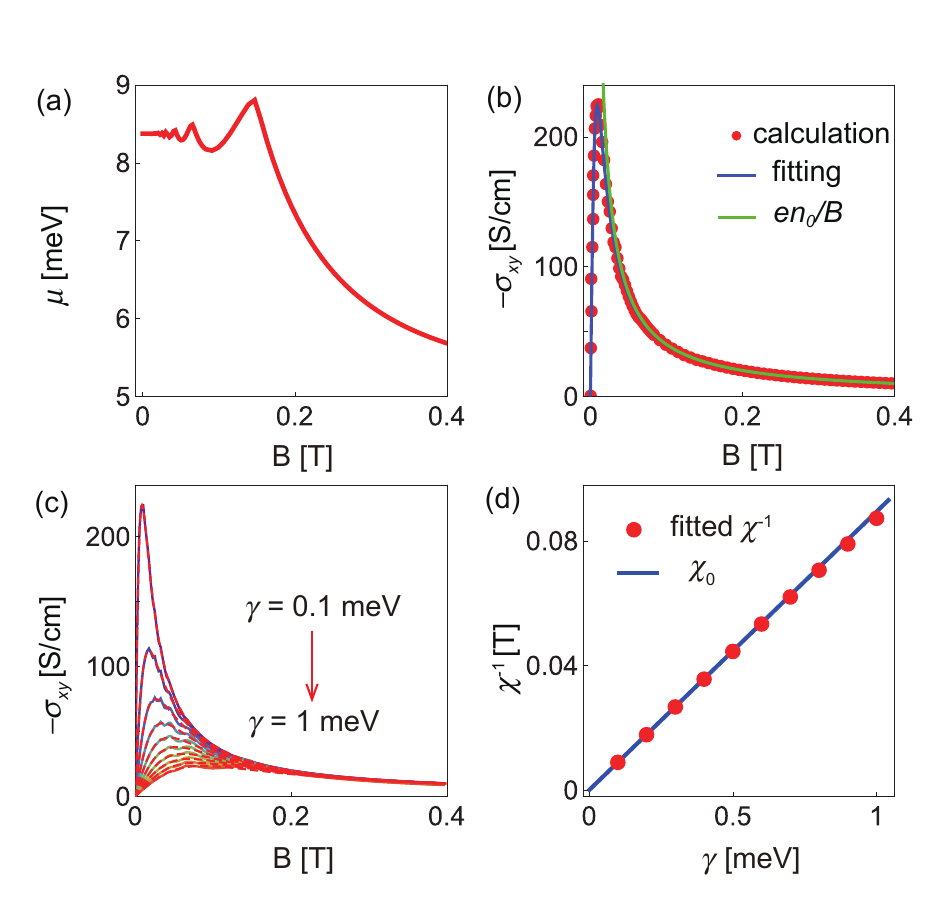}\caption{\label{fig:(a)-By-fixing}(a) By fixing the carrier density as $n_{0}=2.5\times10^{15}\,\mathrm{cm}^{-3}$,
the chemical potential $\mu$ is a function of magnetic field. (b)
Hall conductivity as a function of magnetic field for Zeeman energy
$\omega=\frac{1}{2}g\mu_{B}B$ and constant broadening $\gamma=0.1\,\mathrm{meV}$,
the open circles are the numerical results, blue line is the fitting
curve from the Hall conductivity $\sigma_{xy}=\frac{\chi B\sigma_{D}}{1+\chi^{2}B^{2}}$,
and the green solid line represents the Hall conductivity in the clean
limit. (c) Hall conductivity as a function of magnetic field for several
selected band broadenings, the red dashed lines are the fitting curve
from the Hall conductivity $\sigma_{xy}=\frac{\chi B\sigma_{D}}{1+\chi^{2}B^{2}}$,
and the solid lines are the numerical results from Eq. (\ref{eq:kubo-streda}).
(d)The inverse of fitted mobility as function of band broadenings,
the blue line are the inverse of mobility at $B=0$, $\chi_{0}=\frac{e\hbar}{2\gamma}\frac{v_{x}v_{y}}{\mu}$.}
\end{figure}

For the magnetic-field-dependent Zeeman splitting, i.e., $\omega=\frac{1}{2}g\mu_{B}B$
with $g=20$, it is hard to distinguish the contribution of anomalous
Hall conductivity and conventional orbital Hall conductivity. Setting
a constant broadening width $\gamma=0.1\,\mathrm{meV}$ and carrier
density $n_{0}=2.5\times10^{15}\,\mathrm{cm^{-3}}$, following the
same procedure, we can calculate the Hall conductivity with disorder
effect, which has been given in Fig. \ref{fig:(a)-By-fixing}. When
the carrier density is fixed, as shown in Fig. \ref{fig:(a)-By-fixing}(a),
the chemical potential varies as a function of magnetic field, and
it decreases monotonically in the strong magnetic field. Besides,
the Hall conductivity can be described by the orbital Hall conductivity
$\sigma_{xy}=\frac{\chi B\sigma_{D}}{1+\chi^{2}B^{2}}$ very well,
as indicated by the blue line in Fig. \ref{fig:(a)-By-fixing}(b).
Similar to the constant spin-splitting case, the Hall conductivities
with disorder effect coincide with the Hall conductivity in the clean
limit {[}$\sigma_{xy}=-en_{0}/B$, the Green line in Fig. \ref{fig:(a)-By-fixing}(b){]}
in the high field regime. Besides, we also calculate the Hall conductivity
for several different band broadenings in Fig. \ref{fig:(a)-By-fixing}(c),
where the dip position shifts to the high magnetic field by increasing
$\gamma$. The background of total Hall conductivity {[}solid lines{]}
can be well-fitted by $\sigma_{xy}=\frac{\chi B\sigma_{D}}{1+\chi^{2}B^{2}}$
as indicated by the red dashed lines. The obtained mobility $\chi$
{[}red dots{]} has a good agreement with the mobility at zero magnetic
field $\chi_{0}=\frac{e\hbar}{2\gamma}\frac{v_{x}v_{y}}{\mu}$ as
shown in Fig. \ref{fig:(a)-By-fixing}(d), where the Zeeman splitting
is a higher order contribution in magnetic field to the Hall conductivity
and negligible. In Appendix C, we further evaluate the transverse
conductivity to obtain the Hall resistivity, we find the Hall resistivity
is almost linear in magnetic field, which also does not show the signature
of anomalous Hall effect.

Most previous works attribute the anomalous Hall effect to the Berry
curvature effect due to the band degeneracy lifting by the Zeeman
splitting. This effect can be evaluated based on a semiclassical approach
by the integration of the Berry curvature and the magnetic field is
only encoded in the energy level splitting for spin-up and spin-down
electrons. However, in Dirac systems with large spin-orbital coupling
which couples the spin-up and spin-down bands together, the magnetic
field also introduces the vector potential that the canonical momentum
is replaced by the kinetic momentum $\hbar\mathbf{k}\to\hbar\mathbf{k}+e\mathbf{A}$,
leading to the formation of the Landau levels. The semiclassical approach
completely ignores this part of contribution. In the full quantum
mechanical approach here, we treat these two parts of contribution
simultaneously. As previously discussed, The discrepancy between two
approaches becomes more apparent for strong fields especially in the
quantum limit where only the lowest Landau subband is filled and the
semiclassical approach is completely inapplicable. In this regime,
the Hall conductivity decreases as $B^{-1}$ as $B$ increases in
quantum mechanical approach whereas saturates at high fields in semiclassical
approach.

\section{Possible origins}

As the Zeeman effect has been excluded for the anomalous Hall effect,
we expect a new mechanism for it. By summarizing the experiments in
different works, we find that the anomalous Hall effect is more significant
in thin film sample, which is usually several hundred nanometers.
Consider the layer structure of $\mathrm{ZrTe_{5}}$ and small velocity
along the $z$-direction, it can be regarded as a quasi-two dimensional
system, and the localization effect may play important role in the
Hall conductivity as in the pure two-dimensional system. Usually,
the localization effect can be effectively considered by fixed chemical
potential \citep{QHE}. In the clean limit and zero temperature, the
carrier density in Eq. (\ref{eq:carrier density B}) becomes $n_{0}=\frac{k_{F,z}}{2\pi^{2}\ell_{B}^{2}}$
with $k_{F,z}$ the fermi wave vector of lowest Landau level. Then,
plugging $n_{0}$ into Eq. (\ref{eq:Hall_LL}), one obtained the Hall
conductivity in the quantum limit as
\begin{equation}
\sigma_{xy}=-\frac{e^{2}}{2\pi^{2}\hbar}k_{F,z}.\label{eq:Hall_QL}
\end{equation}
It is noted that Eq. (\ref{eq:Hall_QL}) is a general expression for
the Hall conductivity in the quantum limit. Once $k_{F,z}$ is pinned
to a constant due to the localization effect, $\sigma_{xy}$ is quasi-quantized.
For density $n_{0}=\varrho\times10^{15}\mathrm{cm}^{-3}$ at zero
temperature and zero magnetic field, one has $k_{F,z}=(3\pi^{2}n_{0})^{1/3}\frac{(v_{x}v_{y}v_{z})^{1/3}}{v_{z}}$
and $B_{QL}\approx0.314\frac{v_{z}}{(v_{x}v_{y}v_{z})^{1/3}}\varrho^{2/3}\,\mathrm{T}$,
where the system enters the quantum limit for $B>B_{QL}$. In general,
we expect that the critical field for Hall plateau is smaller than
$B_{QL}$ due to the effect of disorder and temperature. This simple
analysis is consistent with the experimental measurements, where the
magnitude of Hall plateau and the corresponding critical field are
increasing functions of carrier density in the low temperatures \citep{Mutch2021arxiv}.

\begin{figure}
\centering{}\includegraphics[width=7.5cm]{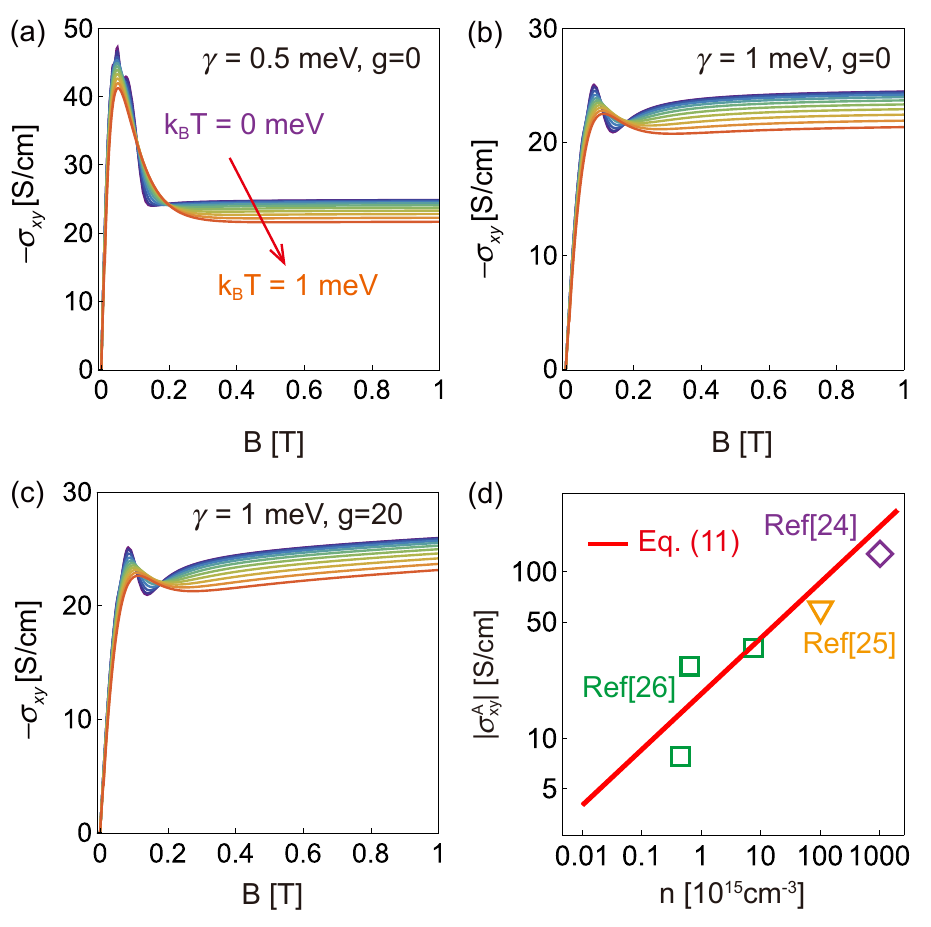}\caption{\label{fig:Hall_fix_mu} Hall conductivity as a function of the magnetic
field for different temperatures. The band broadening is chosen as
$\gamma=0.5\,\mathrm{meV}$ for (a) and (c) and $\gamma=1\,\mathrm{meV}$
for (b). The g-factor is chosen as $g=0$ for (a) and (b) and $g=20$
for (c). The carrier density in the absence of magnetic field is fixed
as $n_{0}=2.5\times10^{15}\,\mathrm{cm^{-3}}$ at different temperatures.
The chemical potential is fixed by varying the magnetic field. (d)
the comparison between Eq. (\ref{eq:Hall_QL}) and experimental data
in literatures.}
\end{figure}

As shown in Fig. \ref{fig:Hall_fix_mu}, by fixing the chemical potential
in the magnetic field, we present the Hall conductivity at different
temperatures. There is a clear quasi-quantized structure in the Hall
conductivity when the system enters the quantum limit regime. The
magnitude of the plateau decreases with increasing temperature, and
is almost independent of the band broadening. The oscillatory part
of Hall conductivity is almost smeared out by a finite temperature.
In addition, the Zeeman effect does not change the results qualitatively,
and it only leads to the upward trend in the high magnetic field as
shown in Fig. \ref{fig:Hall_fix_mu} (c). Moreover, we plot Eq. (\ref{eq:Hall_QL})
and experimental data in literatures \citep{sun2020npj,liu2021PRB,Mutch2021arxiv}
together in Fig. \ref{fig:Hall_fix_mu} (d). Eq. (\ref{eq:Hall_QL})
describes the carrier density dependence of Hall plateau value very
well, which demonstrates that the observed Hall plateau can be attributed
to the fixed chemical potential in magnetic field. Theoretically,
the incommensurate charge density wave could offer one possible mechanism
for the fixed chemical potential in $\mathrm{ZrTe}_{5}$ \citep{tang2019nature,qin2020prl}.
However, the formation of charge density wave also requires the transverse
conductivity vanishes in the corresponding magnetic field regime,
which is inconsistent with the most of experimental measurements for
$\mathrm{ZrTe_{5}}$ samples. Hence, it is anticipated that the fixed
chemical potential is caused by mechanisms other than charge density
wave, such as the localization effect from disorder \citep{Rameshprb1990,Morgensternprb2001,Haudeprl2001}.
Besides, if there is charge transfer between the conduction band and
other strongly scattering additional band, the carrier density in
the conduction band can generically vary with field \cite{Gooth2022rpp}.
Correspondingly, the fermi wave vector $k_{F,z}$ might be field insensitive
and is approximately a constant. The theoretical mechanism behind
these scenarios requires further study in the future.

\section{Summary and discussion}

In summary, we have studied the Hall conductivity for $\mathrm{ZrTe}_{5}$
and $\mathrm{HfTe_{5}}$ based on the massive Dirac fermions. When
Landau levels are formed in a finite magnetic field, there are two
cases. (i) For a constant spin splitting, $\sigma_{xy}^{A}$ is finite
and robust to the weak disorder at $B=0$, but vanishes in a high
magnetic field and in the clean limit. (ii) For the magnetic field
dependent Zeeman splitting $\omega=\frac{1}{2}\mu_{B}gB$, it is hard
to identify the contribution of anomalous Hall conductivity from the
total Hall conductivity. The Hall resistivity is almost linear in
magnetic field even in the presence of Zeeman effect with a giant
$g$-factor ($g=20$). Actually, the anomalous Hall effect for massive
spin-split Dirac fermions are suppressed by a magnetic field by a
factor $(1+\chi^{2}B^{2})^{-1}$ and vanishes in a finite magnetic
field or in the clean limit as $(1+\chi^{2}B^{2})^{-1}\to0$. Our
calculations indicate that Zeeman field cannot generate the anomalous
Hall effect in $\mathrm{ZrTe_{5}}$ and $\mathrm{HfTe_{5}}$. Even
for constant Zeeman splitting, the anomalous Hall effect is suppressed
in the strong magnetic field, and the calculation from the semiclassical
treatment cannot be simply extended to the strong magnetic field.
If the chemical potential is fixed in the magnetic field, there is
a plateau in Hall conductivity, which might provide an explanation
for the observed anomalous Hall effect in experiments.

\section*{Acknowledgments}

We thank Di Xiao and Jiun-Haw Chu for helpful discussions. This work
was supported by the National Key R\&D Program of China under Grant
No. 2019YFA0308603; the Research Grants Council, University Grants
Committee, Hong Kong under Grant No. C7012-21G and No. 17301220; the
Scientific Research Starting Foundation of University of Electronic
Science and Technology of China under Grant No. Y030232059002011;
and the International Postdoctoral Exchange Fellowship Program under
Grant No. YJ20220059.

\appendix

\section{Anomalous Hall Conductivity without Landau level}

In this section, we simply consider the case semi-classically, where
the effect of magnetic field can be encoded into the Zeeman energy
as $\omega=g_{z}\mu_{B}B/2$ with $g_{z}=21.3$ the g-factor and $\mu_{B}=5.788\times10^{-2}\,\mathrm{meV\cdot T^{-1}}$
the Bohr magneton \cite{liu2016nc}. Then, the low energy Hamiltonian
in Eq. (\ref{eq:massive dirac}) becomes
\begin{equation}
H(k)=m\tau_{z}+\omega\sigma_{z}+v_{i}\hbar k_{i}\Gamma_{i}.\label{eq:massive dirac-1}
\end{equation}
Solving the eigen equation, $H|\psi\rangle=\varepsilon|\psi\rangle$,
we can find the energy spectrum as
\[
\varepsilon_{s\zeta}=\zeta\sqrt{\left(m_{\perp}+s\omega\right)^{2}+\left(v_{z}\hbar k_{z}\right)^{2}},
\]
where $m_{\perp}=\sqrt{m^{2}+\hbar^{2}\left(v_{x}^{2}k_{x}^{2}+v_{y}^{2}k_{y}^{2}\right)}$
, $s=\pm$ represents two splitting states because of the Zeeman effect,
$\zeta=+$ is for the conduction band and $\zeta=-$ is for the valence
band. The system becomes a nodal line semimetal when $\omega>m$,
and the nodal ring is given by $\hbar^{2}\left(v_{x}^{2}k_{x}^{2}+v_{y}^{2}k_{y}^{2}\right)=\omega^{2}-m^{2}$
and $k_{z}=0$.

The corresponding eigenstates are found as 
\begin{align*}
|\psi_{\lambda}\rangle=\begin{pmatrix}\zeta\cos\frac{\phi_{s\zeta}}{2}\cos\frac{\theta_{s}}{2}\\
s\mathrm{sign}(k_{z})\sin\frac{\phi_{s\zeta}}{2}\sin\frac{\theta_{s}}{2}e^{i\phi_{k}}\\
s\zeta\cos\frac{\phi_{s\zeta}}{2}\sin\frac{\theta_{s}}{2}e^{i\phi_{k}}\\
\mathrm{sign}(k_{z})\sin\frac{\phi_{s\zeta}}{2}\cos\frac{\theta_{s}}{2}
\end{pmatrix} & ,
\end{align*}
where the angles $\phi_{s\zeta}$, $\theta_{s}$ and $\phi_{k}$ are
defined as $\cos\phi_{s\zeta}=\frac{\omega+sm_{\perp}}{\varepsilon_{s\zeta}}$,
$\cos\theta_{s}=\frac{sm}{m_{\perp}}$ and $e^{i\phi_{k}}=\frac{v_{x}k_{x}+iv_{y}k_{y}}{\sqrt{v_{x}^{2}k_{x}^{2}+v_{y}^{2}k_{y}^{2}}}$.
The subscript $\lambda$ denotes the quantum number $s$ and $\zeta$.

At the zero magnetic field, the anomalous Hall conductivity can be
attributed to the nonzero Berry curvature of band structure as \cite{xiao2010rmp,nagaosa2010rmp}
\begin{equation}
\sigma_{xy}=\frac{e^{2}}{V\hbar}\sum_{k,\lambda}\Omega_{z}^{\lambda}n_{F}(\varepsilon_{\lambda}-\mu),\label{eq:AHE_Berry-1}
\end{equation}
where $\Omega_{\ell}^{\lambda}$ is the $\ell$th component of Berry
curvature vector of the $\lambda$th band. For well-separated bands,
$\Omega_{\ell}^{\lambda}$ can be expressed as 
\[
\Omega_{\ell}^{\lambda}=\hbar^{2}\epsilon_{ij\ell}\sum_{\lambda^{\prime}\ne\lambda}\frac{\mathrm{Im}[v_{\lambda\lambda^{\prime}}^{i}v_{\lambda^{\prime}\lambda}^{j}]}{(\varepsilon_{\lambda}-\varepsilon_{\lambda^{\prime}})^{2}},
\]
where $\epsilon_{ij\ell}$ is the Levi-Civita antisymmetric tensor
with $i,j,\ell$ standing for $x,y,z$. $v_{\lambda\lambda^{\prime}}^{i}=\langle\psi_{\lambda}|\hat{v}^{i}|\psi_{\lambda^{\prime}}\rangle$
is the matrix element of velocity operator $\hat{v}^{i}$ in the eigen
basis. For the massive Dirac fermions with Zeeman splitting, we can
evaluate the $z$ component of Berry curvature as $\Omega_{z}^{\zeta s}=\frac{smv_{x}v_{y}\hbar^{2}}{2m_{\perp}^{3}}.$
Here $\Omega_{z}^{\zeta s}$ is independent of band index $\zeta$
and momentum $k_{z}$, and its sign depends on the band index $s$
and Dirac mass $m$. The magnitude of $\Omega_{z}^{\zeta s}$ is a
decreasing function of $k_{\perp}$ and has a maximum at $k_{\perp}=0$
as $|\Omega_{z}^{\zeta s}(k_{\perp}=0)|=\frac{\hbar^{2}v_{x}v_{y}}{2m^{2}}$,
and it vanishes as $k_{\perp}\to+\infty$. Then, we arrive the Hall
conductivity as 
\begin{align}
\sigma_{xy}= & e^{2}v_{x}v_{y}\hbar\sum_{s\zeta}\int\frac{d^{3}k}{(2\pi)^{3}}\frac{sm}{2m_{\perp}^{3}}n_{F}(\varepsilon_{s\zeta}-\mu)\label{eq:AHE-1}
\end{align}
It is easy to check that $\sigma_{xy}(-\mu,\omega)=-\sigma_{xy}(\mu,\omega)$
and $\sigma_{xy}(-\omega,\mu)=-\sigma_{xy}(\omega,\mu)$; hence, the
anomalous Hall effect is asymmetric about the chemical potential and
Zeeman energy. When the chemical potential is inside the band gap
and temperature is zero, $n_{F}(\epsilon_{+\zeta}-\mu)=n_{F}(\epsilon_{-\zeta}-\mu)$,
$\sigma_{xy}=0$; otherwise, $\sigma_{xy}(\omega\ne0,\mu)\ne0$. Besides,
as $\varepsilon_{s\zeta}$ and $m_{\perp}$ are even in $m$, $\sigma_{xy}$
is odd in Dirac mass $m$ and vanishes when $m=0$. The finite Dirac
mass is essential for the presence of anomalous Hall effect in $\mathrm{ZrTe}_{5}$.
For simplicity, we put the chemical potential inside the conduction
band $(\mu>0)$, and consider $\omega\ge0$ and $m>0$ in the following
discussion. At the zero temperature, Eq. (\ref{eq:AHE-1}) can be
further simplified as 
\begin{align}
\sigma_{xy}= & \frac{m}{2\pi\hbar v_{z}}\frac{e^{2}}{h}\sum_{s}\int_{|m|+s\omega}^{+\infty}dt\frac{s\sqrt{\mu^{2}-t^{2}}}{(t^{2}-s\omega)^{2}}\theta(\mu^{2}-t^{2})\label{eq:AHE_T_0-1}
\end{align}
where $\theta(x)$ is the unit-step function. We can define sum of
the two integrals as $J$; then, $\sigma_{xy}=\frac{m}{2\pi\hbar v_{z}}\frac{e^{2}}{h}J$.

\begin{figure}
\centering{}\includegraphics[width=8cm]{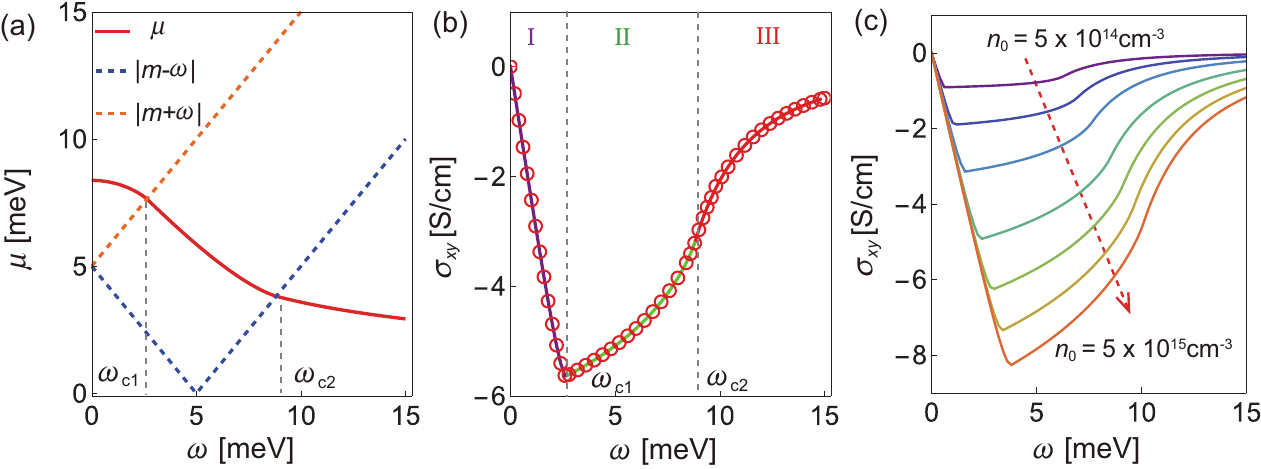}\caption{\label{fig:AHE_omega-1}(a)Chemical potential and (b) Hall conductivity
as a function of the Zeeman energy for a given carrier density. In
(a), the dashed lines are the energy of the band edge of conduction
bands at $k=0$. In (b), red open circles are numerically calculated
from Eq. (\ref{eq:AHE-1}), the purple line is calculated by Eq. (\ref{eq:AHE_R1-1}),
the green line is calculated by Eq. (\ref{eq:AHE_R2-1}), and the
red line is calculated by Eq. (\ref{eq:AHE_R3-1}). (c)The Hall conductivity
as a function of the Zeeman energy for different given carrier density.
The calculating parameters are chosen as $m=5\,\mathrm{meV}$.}
\end{figure}

If we fix carrier density as a constant, $\mu$ is a function of $\omega$
and can be solved from the following equation, 
\begin{align*}
n_{0}= & \sum_{s,\chi=\pm}\int\frac{d^{3}k}{(2\pi)^{3}}\chi n_{F}(\varepsilon_{s}-\chi\mu).
\end{align*}
For instance, we set $n_{0}=2.5\times10^{15}\,\mathrm{cm^{-3}}$,
the obtained chemical potential $\mu$ decreases with the increasing
of $\omega$. There are two critical Zeeman energy, $\omega_{c1}$
and $\omega_{c2}$. As shown in Fig. \ref{fig:AHE_omega-1}(a), when
$0<\omega<\omega_{1c},$ $\mu$ intersect with both bands of $s=+$
and $s=-$. When $\omega>\omega_{c1}$, $\mu$ intersects with band
$s=-$ only. If $\omega$ is further larger than $\omega_{2}$, $\mu$
is lower than the band edge of $s=-$ at $k=0$.

Accordingly, there are three regimes for the nonzero $J$. When $\omega$
is smaller than $\omega_{c1}$,
\begin{align}
J= & \sum_{s}J_{s}(\omega,\mu)\label{eq:AHE_R1-1}
\end{align}
where
\begin{align*}
J_{s}(\omega,\mu)= & \frac{\omega}{\mu_{\omega}}\ln\left(\frac{\mu m}{\mu_{\omega}(\mu_{s}+\mu_{\omega})-s\omega m}\right)\\
 & +\frac{s\mu_{s}}{m}-s\cos^{-1}\left(\frac{m+s\omega}{\mu}\right)
\end{align*}
with $\mu_{s}=\sqrt{\mu^{2}-(s\omega+m)^{2}}$ and $\mu_{\omega}=\sqrt{\mu^{2}-\omega^{2}}$.
As indicated by the regime I in Fig. \ref{fig:AHE_omega-1}(a), the
anomalous Hall conductivity is an decreasing function of $\omega$.

If we further increase the Zeeman energy so that $\omega_{c1}<\omega<\omega_{c2}$,
the system enters the regime II in Fig. \ref{fig:AHE_omega-1}(a),
the dimensionless coefficient $J$ becomes 
\begin{align}
J= & J_{-}(\omega,\mu)\label{eq:AHE_R2-1}
\end{align}
which is a increasing function of $\omega$.

If the Zeeman energy is so large that $\omega\ge\omega_{c2}$, as
shown by the regime III in Fig. \ref{fig:AHE_omega-1}(a), the coefficient
$J$ is found as

\begin{align}
J= & -\pi\left(\frac{\omega}{\sqrt{\omega^{2}-\mu^{2}}}-1\right)\label{eq:AHE_R3-1}
\end{align}
which increases with the increasing of $\omega$ and approaches zero
as $\omega\gg\mu$.

It is noted that $\sigma_{xy}$ reaches its max value when $\omega=\omega_{c1}$,
and the corresponding maximum value are
\begin{align*}
\sigma_{xy}^{max}= & \frac{m}{2\pi\hbar v_{z}}\frac{e^{2}}{h}J_{-}(\omega_{c1},\omega_{c1}+m)
\end{align*}
For $m=5\,\mathrm{meV}$, $n_{0}=2.5\times10^{15}\mathrm{cm^{-3}}$,
$\omega_{c1}=2.65\,\mathrm{meV}$, $\frac{m}{2\pi\hbar v_{z}}\frac{e^{2}}{h}\approx9.25\,\Omega^{-1}\cdot\mathrm{cm^{-1}}$,
and $|\sigma_{xy}^{max}|\approx5.63\,\Omega^{-1}\cdot\mathrm{cm^{-1}}$.
Furthermore, $J_{-}(\omega_{c1},\omega_{c1}+m)$ can be written as
a decreasing function of $\omega_{c}/m$. There are several ways to
enlarge the magnitude of $\sigma_{xy}^{max}$. On the one hand, we
can reduce the fermi velocity $v_{z}$, on the other hand, we can
increase the carrier density so that $\omega_{c}$ can be enhanced
as shown in Fig. \ref{fig:AHE_omega-1}(c). Besides, if we keep the
ratio $\omega_{c}/m$ as a constant, $|\sigma_{xy}^{max}|$ will also
increase with the increasing of Dirac mass $m$.

\section{Landau level}

In a finite perpendicular magnetic field $B$, in terms of the ladder
operators $a$ and $a^{\dagger}$, the Hamiltonian in Eq. (\ref{eq:massive dirac})
can be expressed as
\[
H=\begin{pmatrix}m+\omega & 0 & \eta a & v_{z}\hbar k_{z}\\
0 & m-\omega & v_{z}\hbar k_{z} & -\eta a^{\dagger}\\
\eta a^{\dagger} & v_{z}\hbar k_{z} & -m+\omega & 0\\
v_{z}\hbar k_{z} & -\eta a & 0 & -m-\omega
\end{pmatrix}.
\]
For $n\ge1$, using the ansatz $|\psi_{\lambda}\rangle=[c_{\lambda1}|n-1\rangle,c_{\lambda2}|n\rangle,c_{3\lambda}|n\rangle,c_{4\lambda}|n-1\rangle]^{T}$
with $n=1,2,3,\cdots$, we can solve out the eigen spectrum from the
eigen equation $H|\psi_{\lambda}\rangle=\varepsilon_{\lambda}|\psi_{\lambda}\rangle$
as

\[
\varepsilon_{n\zeta s}=\zeta\sqrt{(E_{n}+s\omega)^{2}+v_{z}^{2}\hbar^{2}k_{z}^{2}},
\]
where $s=\pm$ represents two splitting states because of the Zeeman
effect, $\zeta=+$ is for the conduction band and $\zeta=-$ is for
the valence band, $E_{n}=\sqrt{m^{2}+n\eta^{2}}$. The corresponding
eigen states are given by
\begin{align*}
|\psi_{\lambda}\rangle=\begin{pmatrix}\zeta\cos\frac{\phi_{ns\zeta}}{2}\cos\frac{\theta_{ns}}{2}|n-1\rangle\\
s\mathrm{sign}(k_{z})\sin\frac{\phi_{ns\zeta}}{2}\sin\frac{\theta_{ns}}{2}|n\rangle\\
s\zeta\cos\frac{\phi_{ns\zeta}}{2}\sin\frac{\theta_{ns}}{2}|n\rangle\\
\mathrm{sign}(k_{z})\sin\frac{\phi_{ns\zeta}}{2}\cos\frac{\theta_{ns}}{2}|n-1\rangle
\end{pmatrix} & ,
\end{align*}
where $\cos\phi_{ns\zeta}=\frac{\omega+sE_{n}}{\varepsilon_{ns\zeta}}$
and $\cos\theta_{ns}=\frac{sm}{E_{n}}$. The subscript $\lambda$
denotes the quantum number $n,s,\zeta$.

When $n=0$, we can find the eigen energy and eigen states as 
\[
\varepsilon_{0\zeta}=\zeta\sqrt{v_{z}^{2}\hbar^{2}k_{z}^{2}+(m-\omega)^{2}},
\]
\[
|\psi_{0\zeta}\rangle=\begin{pmatrix}0\\
\mathrm{sign}(k_{z})\sin\frac{\phi_{0\zeta}}{2}|0\rangle\\
\zeta\cos\frac{\phi_{0\zeta}}{2}|0\rangle\\
0
\end{pmatrix},
\]
where $\cos\phi_{0\zeta}=\frac{\omega-m}{\varepsilon_{0\zeta}}$.

In the Landau level basis, the matrix element of velocity operator
$v_{\lambda\lambda^{\prime}}^{i}$ can be evaluated as $v_{\lambda\lambda^{\prime}}^{i}=\langle\psi_{\lambda}|i\hbar^{-1}[H,r_{i}]|\psi_{\lambda^{\prime}}\rangle$.
Along the $x-$ and $y-$ direction, the velocity operators are defined
as $\hat{v}^{x}=i\hbar^{-1}[H,x]=v_{x}\Gamma_{1}$ and $\hat{v}^{y}=i\hbar^{-1}[H,y]=v_{y}\Gamma_{2}$,
respectively. The product of matrix elements of $\hat{v}^{x}$ and
$\hat{v}^{y}$ become,
\begin{align*}
v_{\lambda\lambda^{\prime}}^{x}v_{\lambda^{\prime}\lambda}^{y}= & -i[v_{\lambda\lambda^{\prime}}^{(1)}]^{2}\delta_{n,n^{\prime}-1}+i[v_{\lambda^{\prime}\lambda}^{(1)}]^{2}\delta_{n,n^{\prime}+1}\\
v_{\lambda\lambda^{\prime}}^{x}v_{\lambda^{\prime}\lambda}^{x}= & [v_{\lambda\lambda^{\prime}}^{(2)}]^{2}\delta_{n,n^{\prime}-1}+[v_{\lambda^{\prime}\lambda}^{(2)}]^{2}\delta_{n,n^{\prime}+1}
\end{align*}
where $v_{\lambda\lambda^{\prime}}^{(1)}=\sqrt{v_{x}v_{y}}(c_{\lambda3}c_{\lambda^{\prime}1}-c_{\lambda2}c_{\lambda^{\prime}4})$
and $v_{\lambda\lambda^{\prime}}^{(2)}=v_{x}(c_{\lambda3}c_{\lambda^{\prime}1}-c_{\lambda2}c_{\lambda^{\prime}4})$.
This relation can help us simplify the calculation for the Hall conductivity
under finite magnetic field and temperature.

Besides, $G^{R/A}$ is diagonalized in the Landau level basis, and
the diagonal elements are given by $G_{\lambda}^{R/A}=[\epsilon-\varepsilon_{\lambda}\pm i\gamma]^{-1}$,where
$\lambda=n,s,\zeta$ denote the quantum numbers. By making the integral
by parts for $\epsilon$ in Eq. (\ref{eq:kubo-streda}), $\sigma_{xy}$
becomes 
\begin{align*}
\sigma_{xy}= & \frac{\hbar e^{2}}{\pi^{3}\ell_{B}^{2}}\sum_{\lambda\lambda^{\prime}}\int_{-\infty}^{+\infty}dk_{z}\int_{-\infty}^{+\infty}d\epsilon[v_{\lambda\lambda^{\prime}}^{(1)}]^{2}[-n_{F}^{\prime}(\epsilon-\mu)]\\
\times & \frac{\delta_{n,n^{\prime}-1}}{2(\varepsilon_{\lambda}-\varepsilon_{\lambda^{\prime}})^{2}}\{\tan^{-1}\left(\frac{\varepsilon_{\lambda}-\epsilon}{\gamma}\right)+\tan^{-1}\left(\frac{\epsilon-\varepsilon_{\lambda^{\prime}}}{\gamma}\right)\\
 & -\frac{\gamma(\varepsilon_{\lambda}-\varepsilon_{\lambda^{\prime}})\left((\varepsilon_{\lambda}-\epsilon)(\varepsilon_{\lambda^{\prime}}-\epsilon)+\gamma^{2}\right)}{\left((\varepsilon_{\lambda}-\epsilon)^{2}+\gamma^{2}\right)\left((\varepsilon_{\lambda^{\prime}}-\epsilon)^{2}+\gamma^{2}\right)}\}
\end{align*}
which is only contributed from the states near the fermi surface at
the low temperature. In the weak scattering limit ($\gamma\to0$),
the terms in the big parentheses becomes $\frac{\pi}{2}[\mathrm{sgn}(\varepsilon_{\lambda}-\epsilon)+\mathrm{sgn}(\epsilon-\varepsilon_{\lambda^{\prime}})]$.
After performing the integral of $\epsilon$, one arrives Eq. (\ref{eq:Hall-clean}).

\section{Transverse conductivity and resistivity}

\begin{figure}
\begin{centering}
\includegraphics[width=8cm]{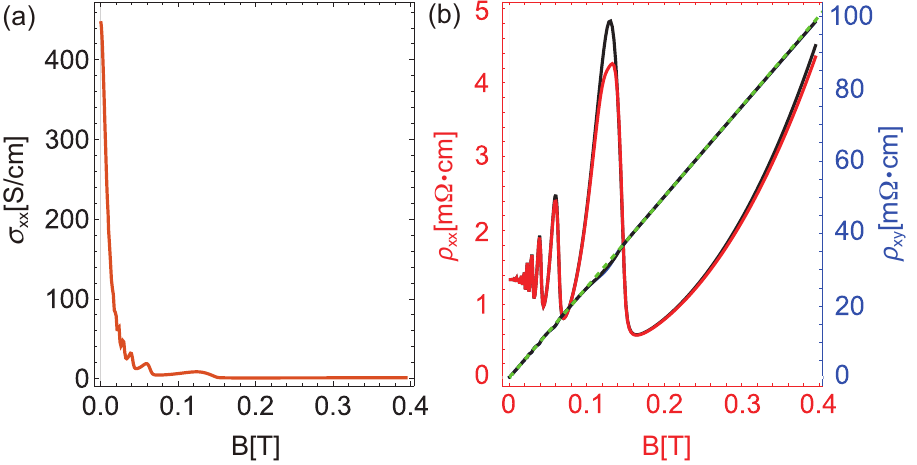}
\par\end{centering}
\centering{}\caption{(a)\label{fig:(a)Transverse-conductivity-as}Transverse conductivity
as a function of magnetic field. (b)The resistivity (red and black
line), Hall resistivity (blue line), and the linear background $B/n_{0}e$
(green line). For the resistivity, the $g$ factors are set as $g=20$
and $g=0$ for red and black line, respectively. For the Hall resistivity,
the results of $g=20$ and $g=0$ are almost overlapped with each
other, here we only present the result for $g=20$.}
\end{figure}

As the magnitude of anomalous Hall conductivity is much smaller than
the orbital Hall conductivity for $\omega=\frac{1}{2}g\mu_{B}B$,
it is hard to see the anomalous contribution. To further confirm our
conclusion in last part, we calculate the Hall resistivity to see
wether there is a nonlinear behavior in the Hall curve or not. To
obtain the elements of resistivity matrix, we need to further calculate
the transverse conductivity $\sigma_{xx}$. According to the Kubo-Strda
formula, the transverse conductivity $\sigma_{xx}$ in the Landau
level basis is given by \cite{Mahan,Streda1982,wang2018prb}
\begin{align*}
\sigma_{xx}= & \frac{e^{2}\hbar}{2\pi^{3}\ell_{B}^{2}}\sum_{\lambda\lambda^{\prime}}\int_{-\infty}^{+\infty}dk_{z}[v_{\lambda\lambda^{\prime}}^{(2)}]^{2}\delta_{n,n^{\prime}-1}\\
 & \int_{-\infty}^{\infty}[-n_{F}^{\prime}(\epsilon-\mu)]\text{Im}G_{\lambda}^{R}\text{Im}G_{\lambda^{\prime}}^{R}d\epsilon
\end{align*}
Setting the calculation parameter identical to the one in main-text,
we obtain the transverse conductivity as shown in Fig. \ref{fig:(a)Transverse-conductivity-as}(a),
where the transverse conductivity decays quickly with the increasing
of magnetic field and display an oscillating behavior for the moderate
strong magnetic field. Besides, the transverse conductivity along
$y$ direction can be obtained as $\sigma_{yy}=(\frac{v_{y}}{v_{x}})^{2}\sigma_{xx}$.

Taking advantage of the obtained Hall conductivity and transverse
conductivity, we can derive the transverse and Hall resistivity as
\begin{align*}
\rho_{xx} & =\frac{\sigma_{yy}}{\sigma_{xx}\sigma_{yy}+\sigma_{xy}^{2}},\\
\rho_{xy} & =-\frac{\sigma_{xy}}{\sigma_{xx}\sigma_{yy}+\sigma_{xy}^{2}}.
\end{align*}
As shown in Fig. \ref{fig:(a)Transverse-conductivity-as}(b), there
are quantum oscillations in $\rho_{xx}$ and $\rho_{xy}$, and the
oscillations split into two components in high magnetic field due
to the Zeeman energy. The background of Hall resistivity is almost
linear in magnetic field as $\rho_{xy}=\frac{B}{n_{0}e}$, which means
there is no anomalous Hall effect due to the Zeeman energy. Besides,
although the large transverse magneto-conductivity, there is almost
no transverse magneto-resistivity. Hence, despite the Zeeman effect
breaks the time-reversal symmetry, there is no linear magneto-resistivity
in the weak magnetic field along the transverse configuration. For
comparison, we also present the resistivity without Zeeman energy
($g=0$) as indicated by the black line in Fig. \ref{fig:(a)Transverse-conductivity-as}(b).
There is no qualitative difference between the cases of $g=0$ and
$g=20$. The Hall resistivity is also linear in magnetic field for
both $g=0$ and $g=20$.

\end{document}